\documentclass[epjST]{svjour}
\usepackage{graphics}

\usepackage{amsfonts,amsmath}

\usepackage{bm} \usepackage{bbm} \usepackage{dcolumn}
\usepackage{epsfig} \usepackage{latexsym}

\usepackage{subfig}

\begin{document}

\title{Impact of surface reflection on transmission eigenvalue statistics and energy distributions inside random media}
\author{Xiaojun Cheng\inst{1} \and Chushun Tian\inst{2,3} \fnmsep\thanks{\email{ctian@mail.tsinghua.edu.cn}} \and Zachary Lowell\inst{1} \and Liyi Zhao\inst{2} \and Azriel Z. Genack\inst{1}\fnmsep\thanks{\email{genack@qc.edu}}}

\institute{Department of Physics, Queens College and Graduate Center, City University of New York, Queens, New York 11367, USA \and Institute for Advanced Study, Tsinghua University, Beijing 100084, China \and PhLAM - Laboratoire de Physique des Lasers, Atomes et Mol$\acute{\rm e}$cules, Universit$\acute{\rm e}$ de Lille 1, F-59655  Villeneuve d'Ascq, France}
\abstract{  
The impact of surface reflection upon transmission through and energy distributions within random media has generally been described in terms of the boundary extrapolation lengths $z_b, z_b'$ at the input and output end of an open sample, which are the distance beyond the sample surfaces at which the energy density within the sample extrapolates to zero \cite{Morse,Lagendijk89,Pine91,Genack93}. The importance of reflection at the sample boundaries plays a key role in the scaling of transmission \cite{1987a,1997d}. Here we consider the impact of surface reflection on the propagation of diffusive waves \cite{Cheng13,Tian15} in terms of the modification of the distribution of transmission eigenvalues (DTE) \cite{Dorokhov84,Imry86,Mello88,Nazarov94,Rossum95,Kogan95,Beenakker97,Rossum99}. We review our finding of a transition in the analytical form of the DTE at the point that the sample length equals $|z_b-z_b'|$. The highest transmission eigenvalue for stronger asymmetry in boundary reflection is strictly smaller than unity. The average transmission and profiles of energy density inside the sample can still be described in terms of the sample length, $L$, and the boundary extrapolation lengths on both sides of the sample, $z_b, z_b'$. For localized waves, we find the energy density profile within the sample is a segment of the distribution that would be found in a longer sample with length $L+z_b+z_b'$. These results suggest new ways of controlling wave interference in both diffusive and localized systems by varying boundary reflectivity.
} 
\maketitle

\section{Introduction}
\label{intro}
The transmission matrix provides a powerful approach to calculating the statistics of electrical conductance and optical transmission \cite{Dorokhov84,Imry86,Mello88,Nazarov94,Rossum95,Kogan95,Beenakker97,Rossum99}. The nature of transport depends upon the proximity to the localization transition \cite{Abrahams79} which originates from the coherent backscattering of waves \cite{1985g,1985h,Maynard86,Maynard88,Altshuler91,Akkermans07}. This is given in terms of the dimensionless conductance, $g$, which is the average electrical conductance in units of the quantum of conductance, $e^2/h$, and is equivalent to the optical transmittance $T$, $g=\langle T\rangle$, where $\langle...\rangle$ represents the average over an ensemble of statistically equivalent samples \cite{Abrahams79,Thouless77,Anderson80,Fisher81}. The transmittance in a single random configuration is the sum of transmission eigenvalues over all independent incident channels,  $T=\sum_n \tau_n$. The methods of random matrix theory apply equally to diffusive and localized waves and have yielded the statistics of transmission in terms of the transmission eigenvalues \cite{Rossum95,Kogan95,Rossum99,Genack14}. For example, the probability distribution of conductance is determined by the correlation between the transmission eigenvalues, which leads to a Gaussian distribution with constant variance for diffusive waves \cite{Imry86,Stone85,1990g}, and a log-normal distribution for localized waves with a variance given by the single parameter scaling hypothesis, $\mathrm{var}(\mathrm{ln} T)= -\langle\mathrm{ln} T\rangle$ \cite{Rossum99,Anderson80,Genack14}. In the crossover to localization, the distribution is found to be a truncated log-normal distribution \cite{Genack14,Marko99,Soukoulis01,Gopar02}. Recently, control of individual incident channels has been demonstrated for light \cite{Mosk08,Popoff10,Choi11,Popoff14}, acoustic waves \cite{Aubry14} and microwave radiations \cite{2012f}. This has spurred interest in the fundamental statistics of the eigenvalues and their use to control transmission and reflection as well as energy density  profiles inside random media \cite{Tian15,Rossum95,Mosk08,Choi11,Genack15,Mosk07,Mosk12,Cao16}.

Random matrix theory has generally neglected the impact of boundary reflectivity. In the absence of boundary reflection, the scaling of average electrical and optical transport in disordered system is characterized by a single parameter, $g$ \cite{Abrahams79,Thouless77}. But boundary reflection is more the rule than the exception; it occurs at the junction between the leads with an electronic device of different material composition and between free space or a medium with uniform index and a medium with higher index of refraction. Boundary reflection can significantly change transmission, reflection, the temporal profile of a transmitted pulse, the probability distributions and correlation functions of transport quantities \cite{Lagendijk89,Pine91,Genack93,Cheng13,Tian15}. 

We compare the DTE, $\rho(\mathcal{T})$, in the presence of boundary reflection to the bimodal distribution $\rho_0(\mathcal{T})=\frac{\xi}{2L}\frac{1}{\mathcal{T}\sqrt{1-\mathcal{T}}}$ \cite{Dorokhov84,Imry86,Dorokhov82}, with $\xi$ and $L$ the localization and sample lengths respectively, obtained in multichannel diffusive media when there is no reflection at the boundary. We briefly review first-principles supersymmetry theory, which allows us to systematically study the impact of surface reflection on key observables characterizing wave transport through random media \cite{Efetov97,Tian13}. We also present new results on the impact of reflection on the average energy density distribution of localized waves within random media.

\section{Supersymmetry field theory for wave transport in open random media}
\label{sec:theory}

\subsection{General structure}
\label{sec:field_theory}

We launch a scalar wave of circular frequency $\omega$ at the input of a multichannel random medium of length $L$, local dimension $d$  and cross-sectional area $A$. The propagation of waves is described by the retarded (advanced) Green's function, defined as $\left(\omega^2_\pm \epsilon({\bf r}) + \nabla^2\right)G^{R,A}_{\omega^2}({\bf r},{\bf
 r}')=\delta({\bf r} -{\bf r}')$. Here $\omega_\pm =
\omega \pm i\delta$ with $\delta$ a positive infinitesimal. The
dielectric function is $\epsilon({\bf r})=1+\delta \epsilon ({\bf r})$: the first term is the air background value, while the second term $\delta \epsilon({\bf r})$ represents spatial fluctuations following Gaussian distributions with no spatial correlations. Dielectric layers are abutted against the left and right open sample boundaries with dielectric constants  $\epsilon^{L,R}$. We set the wave group velocity in air to be unity. 

A generic observable, $O$, can be microscopically expressed in terms of the disorder average of the product of Green's functions. Then supersymmetry field theory is able to cast this expression into a functional integral over the supersymmetric
field $Q(x)$, where $Q \equiv\{Q_{\alpha\alpha'}^{\lambda\lambda'}\}$ is a $4\times 4$ supermatrix, with $\lambda,\lambda'=1,2$ denoting the advanced-retarded (`ar') sectors
representing the different analytic structures of $G^{R,A}$ and $\alpha,\alpha'={\rm f,b}$ the fermionic-bosonic (`fb') sectors. This field obeys the nonlinear constraint, $Q^2=\mathbbm{1}$. For waves in open media, the functional integral has the following general structure (see \cite{Tian13} for a review),
\begin{eqnarray}
\begin{aligned}
O &= \int D[Q]o[Q]e^{-S[Q]},\\
S[Q] &= \frac{\pi\nu A D_0}{4}\int_0^L dx
{\rm str} (\partial_x Q)^2+S_{\rm surface},
\end{aligned}
\label{eq:71}
\end{eqnarray}
where $\nu(\omega)$ is the density of states per unit volume, $D_0=\ell/d$ the Boltzmann diffusion constant,
`str' the supertrace, and $S_{\rm surface}$ the surface action. This expression is valid for both diffusive and localized samples.
It shows that the observable is the average over an `auxiliary microscopic variable', $o[Q]$,
with respect to the weight $e^{-S[Q]}$. The explicit form of $o[Q]$ is observable-dependent.
The first term in the action $S[Q]$ governs spatial fluctuations of $Q$ and
accounts for localization in an infinite multichannel random medium; The second term $S_{\rm surface}$ accounts for the difference between the supersymmetry field theory of infinite \cite{Efetov97} and open \cite{Tian08} media. The explicit form of $S_{\rm surface}$ depends on the variable $O$. We will give its form for different $O$s below. 

\subsection{Transmission eigenvalue statistics}
\label{sec:transmission_eigenvalue_statistics}

The study of transmission eigenvalues was pioneered by Dorokhov \cite{Dorokhov84} and Mello, Pereyra and Kumar \cite{Mello88}. These and later studies did not address the impact of surface reflection, which are of fundamental and practical importance. The impact of surface reflection has been found recently by using supersymmetry field theory \cite{Cheng13,Tian15}. Here we review this formalism.

In the basis of empty waveguide modes, $\phi_{a,b}(y)$, where the indices $a,b$ label the left and right boundaries and $y$ denotes the transverse coordinate, the matrix elements of the transmission matrix, $t_{ba}$, are given by  $t_{ba}=i\sqrt{v_b v_a} \int dy \int dy' \phi_b (y) \phi_a^* (y') G_{\omega^2}^A({\bf r},{\bf r}')$, where $v_a$ is the group velocity of the empty waveguide mode $a$. With the parametrization ${\cal T}\equiv\cosh^{-2}\frac{\phi}{2},\phi>0$, the DTE, $\rho({\cal T})\equiv \left\langle \sum_n \delta ({\cal T}-\tau_n)\right\rangle$, can be expressed as
\begin{eqnarray}
\rho({\cal T}) = \frac{1}{2\pi}[F(\phi+i\pi)+F^*(\phi+i\pi)]
\frac{d\phi}{d{\cal T}}. \label{eq:1}
\end{eqnarray}
The function $F(\phi)$ is the ``observable'' associated with the DTE
\begin{eqnarray}
F(\phi)=-\partial_{\zeta_2}
    \left\langle \frac{{\rm det}(1-\gamma_1\gamma_2
    \hat j \delta_{L}G^{A}_{\omega^2}\hat j \delta_{R}G^{R}_{\omega^2})}
    {{\rm det}(1-\zeta_1\zeta_2
    \hat j \delta_{L}G^{A}_{\omega^2}\hat j \delta_{R}G^{R}_{\omega^2})}\right\rangle\Bigg|_{\theta=i\phi},
\label{eq:80}
\end{eqnarray}
where $\gamma_1=\frac{1}{2}\sin\theta,\, \gamma_2=\tan\frac{\theta}{2}$ ($0<\theta<\pi$), $
\zeta_1=\frac{i}{2}\sinh\phi$ and $\zeta_2=i\tanh\frac{\phi}{2}$. Here,
$\hat j$ is the energy flux operator in the longitudinal direction and $\delta_{L (R)}$ restricts the spatial integral on the left (right) surface.

With the help of Eq.~(\ref{eq:80}), we can express $F(\phi)$ as a functional integral over the supermatrix $Q$
\begin{eqnarray}
F(\phi)
=
-\frac{i\xi}{2}
\int
D[Q]
(Q\partial_x Q)^{21}_{\rm bb} e^{-\frac{\xi}{8}
\int_0^L dx
{\rm str} (\partial_x Q)^2-S_{\rm surface}}\big|_{
x=0,\theta=i\phi},
\label{eq:3}
\end{eqnarray}
with the localization length $\xi=2\pi\nu A D_0$. Note that this conforms to the general structure (\ref{eq:71}). For large channel number, $\omega^{d-1}A\gg 1$, the surface action is 

\begin{eqnarray}
S_{\rm surface}=S_{\rm surface}^L+S_{\rm surface}^R,\nonumber\\
S_{\rm surface}^L=-\frac{\tilde N_d\omega^{d-1}A}{2}\frac{1-R(\epsilon^L)}{1+R(\epsilon^L)}{\rm str} (\Lambda Q(0)),\\
S_{\rm surface}^R=-\frac{\tilde N_d\omega^{d-1}A}{2}\frac{1-R(\epsilon^R)}{1+R(\epsilon^R)}{\rm str} (\Gamma Q(L)).\nonumber
\label{eq:74}
\end{eqnarray}
Here, $\tilde N_d=((4\pi)^{\frac{d-1}{2}}\frac{d-1}{2}\Gamma(\frac{d-1}{2}))^{-1}$, and $\Gamma(x)$ is the Gamma function.
$\epsilon^{i}$ ($i=L,R$) determine $R(\epsilon^{L(R)})$ via
\begin{eqnarray}
\frac{1-R(\epsilon^i)}{1+R(\epsilon^i)}\equiv\left\langle\frac{1-R_{{\bf k}_\perp}(\epsilon^i)}{1+R_{{\bf k}_\perp}(\epsilon^i)}\right\rangle_{{\bf k}_\perp},\quad
\label{eq:81}\\
    R_{{\bf k}_\perp}(\epsilon^i)=\left|\frac{\cos\theta_0-\sqrt{\epsilon^i-\sin^2\theta_0}}{\cos\theta_0+\sqrt{\epsilon^i-\sin^2\theta_0}}\right|^2,\, \theta_0\equiv \arcsin\frac{|{\bf k}_\perp|}{\omega},\label{eq:72}
\end{eqnarray}
with ${\bf k}_\perp$ the label for the empty waveguide modes and $\langle\cdots\rangle_{{\bf k}_\perp}$ the average over all modes for which $|{\bf k}_\perp|\leq \omega$. Note that Eq.~(\ref{eq:72}) is the Fresnel formula for the reflection coefficient. Finally
\begin{eqnarray}
  \Lambda = \left(
                \begin{array}{cc}
                  \mathbbm{1}^{\rm fb} & 0 \\
                  0 & -\mathbbm{1}^{\rm fb} \\
                \end{array}
              \right)^{\rm ar},\quad
\Gamma =
                   \left(
\begin{array}{cc}
                        \cos\theta & -i\sin\theta \\
                        i\sin\theta & -\cos\theta \\
                      \end{array}
                    \right)^{\rm ar}
                    \oplus
\left(
                      \begin{array}{cc}
                        \cosh\phi & \sinh\phi \\
                        -\sinh\phi & -\cosh\phi \\
                      \end{array}
                    \right)^{\rm ar},
                    \label{eq:17}
\end{eqnarray}
are constant supermatrices. This field theory for the DTE is valid for both diffusive ($L\ll \xi$) and localized ($L\gg\xi$) samples.

\subsection{Energy density profiles inside random media}
\label{sec:_intensity}
The transmission eigenvalue statistics fully determine transmission properties, but does not give the energy distribution in the interior of random media. This has been extensively investigated using the supersymmetry field theory \cite{Tian08,Tian10,Tian13a} (see also Ref. \cite{Tian13} for a review). Specically, we consider the spatial correlation function, ${\cal Y} ({\bf r},{\bf r}')$, of energy density at positions ${\bf r}, {\bf r}'$, ${\cal Y} ({\bf r},{\bf r}')  \equiv \left\langle G^A_{\omega^2} ({\bf r},{\bf
 r}') \, G^R_{\omega^2} ({\bf r}',{\bf r})\right\rangle.$ This can be expressed as
\begin{eqnarray}
{\cal Y} ({\bf r},{\bf r}')=\left(\frac{\pi \nu}{\omega}\right)^2
\int D[Q] Q^{12}_{\rm bb}(x)Q^{21}_{\rm bb}(x') e^{-\frac{\xi}{8}
\int_0^L dx
{\rm str} (\partial_x Q)^2-S_{\rm surface}}.
\label{eq:7}
\end{eqnarray}
This quantity does not depend on the transverse coordinate in the present multichannel sample. Thus, from now on, we denote it as ${\cal Y} (x,x')$. Again, it conforms to the general structure (\ref{eq:71}). It differs from Eq.~(\ref{eq:3}) in the pre-exponential factor. In the surface action, the constant supermatrix $\Gamma$ in $S_{\rm surface}^R$ [cf. Eq.~(\ref{eq:74})] is replaced by $\Lambda$ i.e.
\begin{equation}
S_{\rm surface}^R=-\frac{\tilde N_d\omega^{d-1}A}{2}\frac{1-R(\epsilon^R)}{1+R(\epsilon^R)}{\rm str} (\Lambda Q(L)).
\label{eq:8}
\end{equation}

\section{Transmission eigenvalue statistics}
\label{sec:DTE}

In this section we focus on diffusive samples and review the analytical results for $\rho(\mathcal{T})$ obtained from supersymmetry field theory \cite{Cheng13,Tian15}. The basic idea is to apply semiclassical analysis to the field theory. Specifically, from Eqs.~(\ref{eq:3}) and (\ref{eq:74}) we find the saddle point equation, $\partial_x (Q\partial_xQ)=0,$ where $Q$ satisfies the boundary conditions
\begin{eqnarray}
    (2z_b' Q\partial_x Q+[Q,\Lambda])|_{x=0}=0,\qquad (2z_b Q\partial_x Q - [Q,\Gamma])|_{x=L}=0,
    \label{eq:65}
\end{eqnarray}
at the left and right surfaces.
%
From Eq.~(\ref{eq:65}), we see that two new scales, $z_b'$ and $z_b$, appear  
\begin{equation}
\label{eq:79}
    z_b'\equiv\sqrt{\pi}\frac{\Gamma\left(\frac{d+1}{2}\right)}{\Gamma\left(\frac{d}{2}\right)}\frac{\ell}{d}\frac{1+R(\epsilon^L)}{1-R(\epsilon^L)},\quad
    z_b\equiv\sqrt{\pi}\frac{\Gamma\left(\frac{d+1}{2}\right)}{\Gamma\left(\frac{d}{2}\right)}\frac{\ell}{d}\frac{1+R(\epsilon^R)}{1-R(\epsilon^R)}.
\end{equation}
These scales are {\it not} additional parameters in the field theory. Instead, they emerge automatically from the general theory (\ref{eq:3}) and (\ref{eq:74}). As will be seen below, as far as the average conductance (of diffusive samples) is concerned, these two scales are the same as the extrapolation lengths and the effective sample length is $L+z_b'+z_b$ \cite{Lagendijk89,Pine91,Genack93,Rossum99}. 

By solving the saddle point equation with boundary conditions (\ref{eq:65}) and substituting the solution into Eqs. (2) and (4), we obtain $\rho({\cal T})$, which is governed by two dimensionless parameters, $\zeta'\equiv z_b'(\epsilon^L)/L$ and $\zeta\equiv z_b(\epsilon^R)/L$, with $\zeta=\zeta'=0$ corresponding to the origin in Fig. 1. 

\begin{figure}[h]
    \centering
    \includegraphics[width=6.0cm]{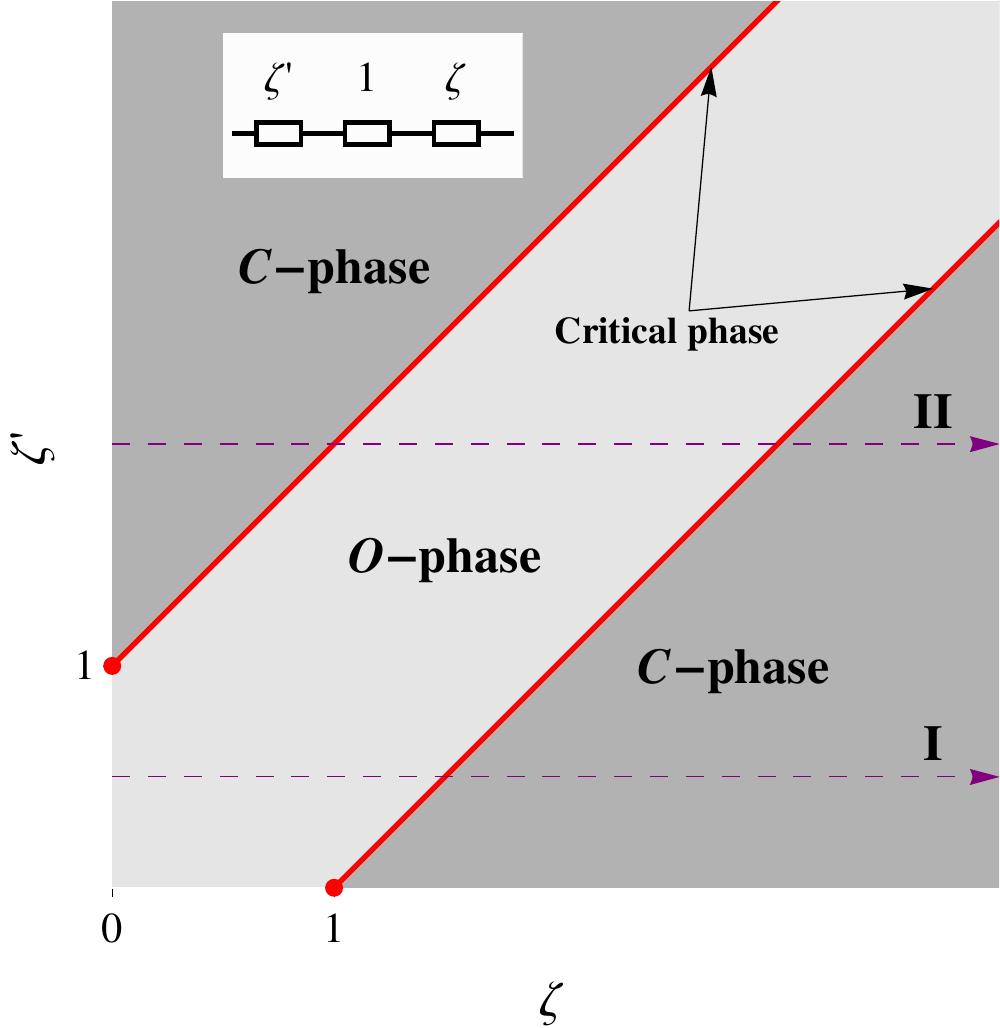}
    \caption{Main panel: the phase diagram of the DTE. Inset: Surface reflection at the input (output) end introduces an surface resistance $\zeta'$ ($\zeta$) (rescaled by the bulk resistance) in series with the bulk resistance.  This figure is from \cite{Tian15}.}
    \label{fig:1}
\end{figure}

Away from the origin, $\rho({\cal T})$ deviates from $\rho_0({\cal T})$ by a factor $ f({\cal T})\equiv \frac{\rho({\cal T})}{\rho_{0}({\cal T})}.$ This is given by
\begin{equation}\label{eq:3f}
    f=(C_{\phi+i\pi}-C_{\phi-i\pi})/(2i\pi)\equiv \Delta C_\phi/(2i\pi),
\end{equation}
where the $C_{\phi\pm i\pi}$ satisfy two closed equations
\begin{eqnarray}
2 \overline{C}_\phi^2 +\Delta C_\phi^2/2 =
\dfrac{\sinh^2 \psi_+}{a \cosh\psi_+ +b} + \dfrac{\sinh^2 \psi_-}{a \cosh\psi_- +b},
\label{eq:4a}
\end{eqnarray}

\begin{eqnarray}
2 \overline{C}_\phi \Delta C_\phi &=&
\dfrac{\sinh^2 \psi_+}{a \cosh\psi_+ +b} - \dfrac{\sinh^2 \psi_-}{a \cosh\psi_- +b},
\label{eq:4b}
\end{eqnarray}
with $\overline{C}_\phi \equiv (C_{\phi+i\pi} + C_{\phi-i\pi})/2$, $\psi_\pm \equiv C_{\phi\pm i\pi}-(\phi\pm i\pi)$, $a = 2\zeta \zeta'$, and $b = \zeta^2 + \zeta'^2$.

\begin{figure}[h]

\centering
\subfloat[] {\includegraphics[width=5.0cm]{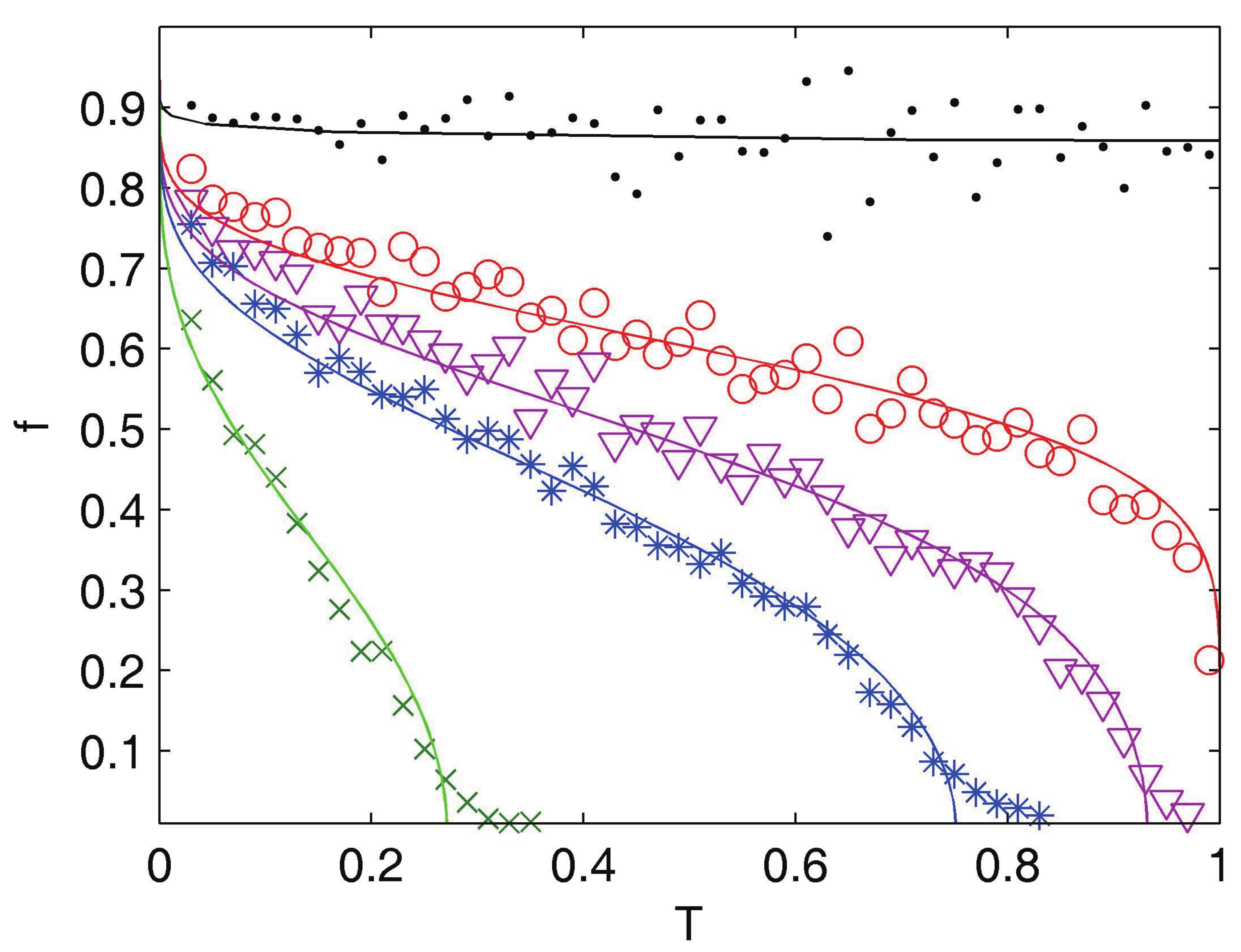}}\qquad
\subfloat[]{\includegraphics[width=5.3cm]{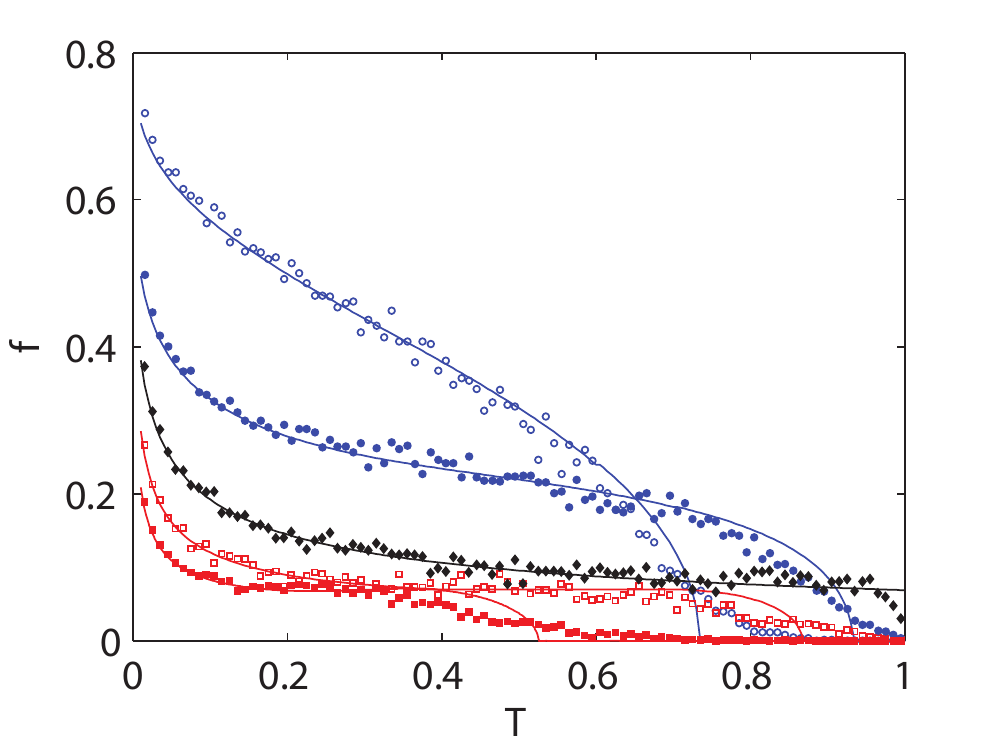}}

    \caption{ (a) Analytic results (solid lines) and simulations show that a single phase transition occurs as the reflection of the output surface increases in diffusive waveguides in which the input surface is perfectly transparent. The dielectric constants on the right surfaces for the curves are $\epsilon^R=1.8, 2, 2.1, 2.2, 2.5$ from top to bottom. This figure is from \cite{Cheng13}. (b) Double phase transitions occur when the input end is sufficiently reflecting, $\epsilon^L=2.1$, and reflectivity on the output increases with dielectric constants of $\epsilon^R=1, 1.9, 2.1, 2.3, 2.4$ from top to bottom near $\mathcal{T}\rightarrow 0$. This figure is from \cite{Tian15}.}
       
\end{figure}

By solving Eqs.~(\ref{eq:4a}) and (\ref{eq:4b}), we obtain a family of curves $f({\cal T})$. Typical results in samples in which surface reflection appears only at a single end of the sample are represented by the solid lines in Fig. 2a. Results for surface in which reflection appears at both ends are shown in Fig. 2b. We see that simulations are consistent with these results .  Figure \ref{fig:1} is the phase diagram in the variables $\zeta,\zeta'$. It is symmetric with respect to the diagonal line $\zeta=\zeta'$ and has three phase regimes characterized by distinct asymptotic behavior of $\rho({\cal T})$ at ${\cal T}\rightarrow 1$. The phase boundaries correspond to the threshold, $|\zeta-\zeta'|=1$, corresponding to $|z_b-z_b'|=L$, in the asymmetry of reflection at the two surfaces. Below this threshold, $\rho({\cal T})$ has the same singularity as $\rho_0({\cal T})$, with $ \rho({\cal T}\rightarrow 1)\sim(1-{\cal T})^{-\frac{1}{2}}$. This implies that perfect transmission is possible and gives the {\it O}-phase regime. Above the threshold, $\rho({\cal T}>{\cal T}_{\rm max})=0$ and the {\it C}-phase regimes follow. At the threshold, the singularity changes to $\rho({\cal T}\rightarrow 1)\sim(1-{\cal T})^{-\frac{1}{3}} $ and perfect transmission can be achieved. This defines the critical phase.


In Fig.~\ref{fig:1}, Line I shows that when reflection at the input surface is small enough that $\zeta'<1$, increasing the reflection at the output (i.e., $\zeta$) leads to a single transition. Tracing along this line, we pass through the critical phase once. Line II shows that when $\zeta'>1$, increasing $\zeta$ leads to a double transition in which the critical phase is passed twice. These results are confirmed by numerical experiments \cite{Cheng13,Tian15}.

To better understand the physical meaning of $\zeta,\zeta'$, we calculate the average transmittance (rescaled by $\xi/L$), $ g\equiv \frac{1}{2}\int_0^1 d{\cal T} {\cal T}\rho({\cal T})=\frac{1}{2}\int_0^1 d{\cal T} \frac{f({\cal T})}{\sqrt{1-{\cal T}}}$, where $f({\cal T})$ is obtained from Eqs.~(\ref{eq:3f}),(\ref{eq:4a}) and (\ref{eq:4b}).
Integrating ${\cal T}$ over $\rho({\cal T})$ 
we obtain $g=1/(1+\zeta+\zeta')$ \cite{Tian15}. Accordingly, $\zeta,\zeta'$ are surface resistances (rescaled by the bulk resistance $L/\xi$). These two resistances and the bulk resistance form a series circuit, as shown in the inset of Fig.~\ref{fig:1}. In other words, the conductance is the same as in a sample without surface reflection but with an effective length of $L+z_b'+ z_b$. This is consistent with the interpretation of $z_b,z_b'$ as boundary extrapolation lengths in studies of the scaling of transmission in diffusive samples \cite{Lagendijk89,Pine91,Genack93,Rossum99}.

\section{Energy profiles inside random media}
\label{sec:_intensity_results}

\subsection{Local diffusion of localized waves in open media: transparent surfaces}
\label{sec:without_edge_reflection}

Supersymmetry field theory for the energy profiles inside open media was first developed in \cite{Tian08}. The perturbative treatments of this theory was performed in \cite{Tian08} and the nonperturbative treatment in \cite{Tian10}. These results were extended to absorbing media in \cite{Tian13a}. In this work, surface reflection was assumed to vanish so that Eq.~(\ref{eq:7}) simplifies to
\begin{eqnarray}
{\cal Y} (x,x')=\left(\frac{\pi \nu}{\omega}\right)^2
\int_b D[Q] Q^{12}_{\rm bb}(x)Q^{21}_{\rm bb}(x') e^{-\frac{\xi}{8}
\int_0^L dx
{\rm str} (\partial_x Q)^2}.
\label{eq:19}
\end{eqnarray}
From this, it was predicted analytically and confirmed numerically that, in open 1D media, ${\cal Y} (x,x')$ for localized waves satisfies a generalized diffusion equation \cite{Tian13,Tian08,Tian10,Tian13a}
\begin{eqnarray}
-\partial_x D(x) \partial_x {\cal Y} (x,x') = \delta(x-x'),
\label{eq:14}
\end{eqnarray}
with the boundary conditions, ${\cal Y} (x,x')|_{x=0} = {\cal Y} (x,x')|_{x=L}=0.$ Equation (\ref{eq:19}) differs from the normal diffusion equation in that the diffusion coefficient is position-dependent, as first proposed in \cite{Wiersma00}. For deeply localized waves, Eq.~(\ref{eq:19}) gives $D(x) \sim e^{-\frac{x(L-x)}{L\xi}}.$ 

%

It was  found in \cite{Tian10} that the function $D(x)$ exhibits novel scaling behavior: it depends on position $x$ via some scaling factor $\lambda(x)$, i.e., $\frac{D(x)}{D_0}=D_\infty(\lambda(x))$, where $D_\infty(\lambda)$ is the scaling function. $\lambda(x)$ is essentially the probability density for a diffusive wave to return to a cross section at depth $x$ in an open random medium. $D_\infty(\lambda)$ is a perturbative expansion of $\lambda$ for $\lambda \ll 1$, which corresponds to weak localization of waves in open random media, and
\begin{equation}\label{eq:25}
    D_\infty(\lambda)\sim e^{-\lambda}, \quad {\rm for}\, \lambda \gg 1,
\end{equation}
corresponds to strong localization. Substituting $\lambda(x)=\frac{x(L-x)}{L\xi}$ into Eq. (18), we recover $D(x)$ given above. It was found in \cite{Tian13a} that this universal scaling holds even for absorbing media
except that $\lambda(x)$ has a more complicated form.


\subsection{Local diffusion of localized waves in open media: reflecting surfaces}
\label{sec:with_edge_reflection}

In the presence of surface reflection, perturbative analysis shows that the surface action has no effect on macroscopic diffusion. However, the explicit form of $D(x)$ does change. We find that $D(x)$ obeys the universal scaling behavior except that the scaling factor changes to $ \lambda(x)=\frac{\left(x+z_b'\right)\left(L+z_b-x\right)}{\xi \left(L+z_b'+z_b\right)}.$ In the localized regime, we obtain
\begin{equation}\label{eq:27}
    D(x)/D_0\propto e^{-\frac{\left(x+z_b'\right)\left(L+z_b-x\right)}{\xi \left(L+z_b+z_b'\right)}}.
\end{equation}
This result is valid for arbitrary $z_b,z_b'$. In addition, we find that the boundary conditions change to
\begin{eqnarray}
\left(z_b'\partial_x-1\right){\cal Y} (x,x')|_{x=0} = \left(z_b\partial_x+1\right){\cal Y} (x,x')|_{x=L}=0.
\label{eq:28}
\end{eqnarray}

To obtain some insights into the physical meanings of $z_b,z_b'$ for localized samples, we solve Eq.~(\ref{eq:14}) with the boundary conditions (\ref{eq:28}) and the local diffusion coefficient (\ref{eq:27}). We introduce the coordinate transformation, $x\rightarrow z(x)$, $dz=\frac{D(x)}{D_0}dx$. This transforms Eq.~(\ref{eq:14}) to
\begin{eqnarray}
-D_0\partial_z^2 {\cal Y} (z,z') = \delta(z-z'),
\label{eq:30}
\end{eqnarray}
and Eq.~(\ref{eq:28}) to
\begin{eqnarray}
\left(z_b'^*\partial_z-1\right){\cal Y} (z,z')|_{z=z(0)} = \left(z_b^*\partial_z+1\right){\cal Y} (z,z')|_{z=z(L)}=0,
\label{eq:32}
\end{eqnarray}
where $z_b'^*=z_b' D_0/D(0)\geq z_b'$ and $z_b^*=z_b D_0/D(L)\geq z_b$. This is the normal diffusion equation in $z$-space with mixed boundary conditions, whose solution is
\begin{equation}\label{eq:31}
    {\cal Y}(z,z')=\frac{1}{D_0}\frac{(z_b'^*+z_<)(L^*+z_b^*-z_>)}{L^*+z_b^*+z_b'^*},
\end{equation}
with $z_{<(>)}={\rm min}({\rm max})\{z,z'\}$ and $L^*=\int_0^L \frac{D(x)}{D_0}dx$. When flux is injected at $x'$ corresponding to $z'$ in the {\it virtual} $z$-space, the intensity drops linearly in the virtual space to zero at a distance of $z_b'^*$ ($z_b^*$) beyond. When localization effects are ignored, $D(x)=D_0$ so that $z_b'^*=z_b'$, $z_b^*=z_b$ and $z(x)=x$. Eq. (\ref{eq:31}) is reduced to the solution to the normal diffusion equation and $z_b',z_b$ acquire the canonical physical meaning of extrapolation lengths \cite{Morse,Lagendijk89,Pine91,Genack93}.

Further insights into the physical meanings of $z_b,z_b'$ are provided by considering the average conductance $g$, which is $\left(\int_0^L \frac{dx}{D(x)}\right)^{-1}$ \cite{Tian10,Wiersma00} in the deeply localized regime ($\xi\ll L$). For simplicity, we take $z_b,z_b'$ to be much smaller than $L$ but it may be larger or smaller than $\xi$. The expression (\ref{eq:27}) for $D(x)$ is then, $g\sim e^{-\frac{L}{4\xi}(1+\zeta+\zeta')}$.

\begin{figure}[h]

\centering
\subfloat[] {\includegraphics[width=5.2cm]{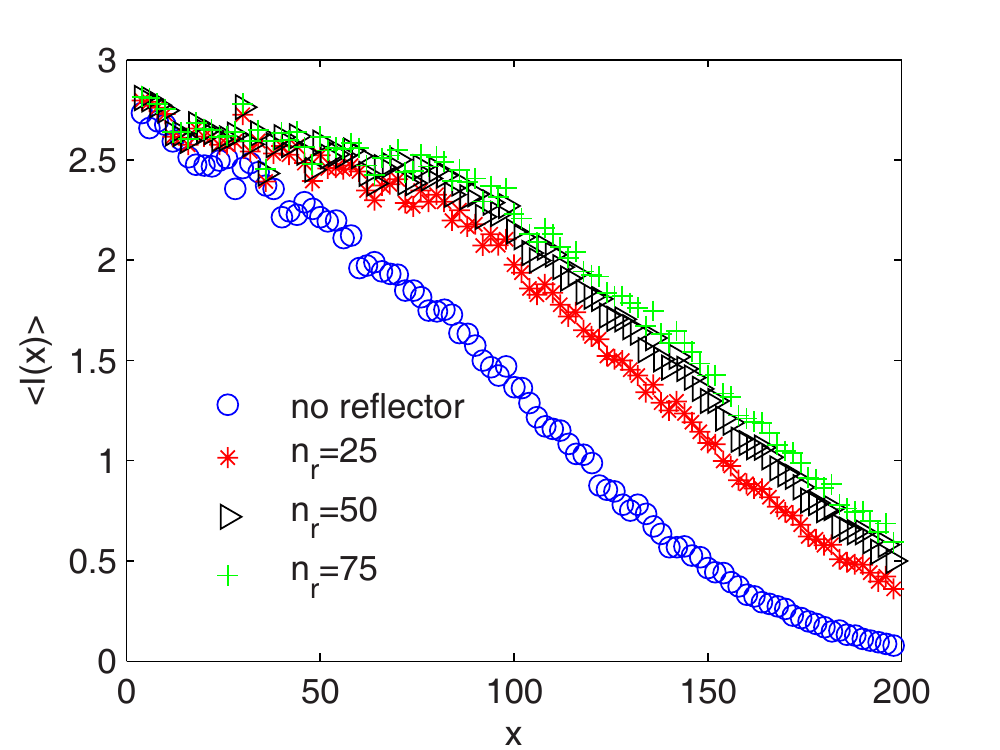}}\qquad
\subfloat[]{\includegraphics[width=5.2cm] {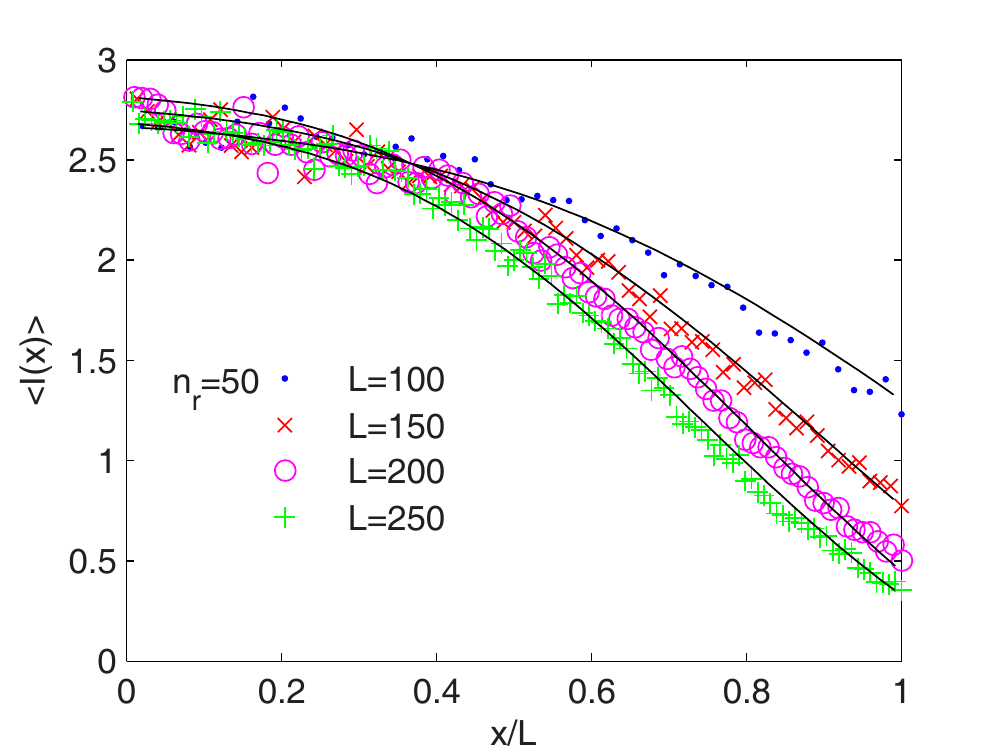}}

    \caption{(a) $\langle I(x)\rangle$ for fixed sample length of $L=200$ layers without boundary reflection at the input for no reflector at the output and the refractive indices of the reflector at the output boundary of $n_r=25, 50, 75$. (b) $\langle I(x)\rangle$ for sample lengths of $100, 150, 200$ and $250$ layers and refractive index of the boundary reflector at the output of $n_r=50$. The solid black lines are predictions from Eq. (19). The parameter $z_b$ is $120$ layers for all the samples.}
       
\end{figure}

We carried out $1$D simulations using the scattering matrix method to explore the impact of surface reflection on the energy density profiles of localized waves inside random media. The sample is composed with binary layers of refractive indices of $n_1=1$  and  $n_2=1.6$. The average thickness of each layer is $1$. The thickness of the material with high refractive index  $n_2$ is fixed and the thickness of the $n_1$  layers varies randomly between $0.5$ and $1.5$. Waves are launched from the left and the incident wavelength is between $1.712$ and $1.760$. A layer with refractive index $n_r=25, 50, 75$ and thickness $2$ is placed over the output of the sample. The average for $500$ configurations of energy density profiles for different boundary reflectors of the same sample length of $200$ layers are shown in Fig. 3a. We see that by varying the boundary reflectivity, the intensity profiles are no longer symmetric with respect to the center of the sample. Instead, they are a truncated profile of a longer sample. To obtain the value of $z_b$, we fix the refractive index of the reflector at the output to be $n_r=50$ and obtain the profile of energy density in samples of length $100, 150, 200, 250$ layers.  We find that the value of $z_b$ at the output surface has the same value of $120$ layers for all sample lengths. The simulation results are in excellent agreement with analytical predictions given by Eq. (19), with $z_b'$ set to zero, as shown in Fig. 3b. These results indicate that intensity profiles within localized samples can be viewed as a segment of the profile of a longer sample with effective sample length of $L^*$ in virtual z-space. Correspondingly, $z_b'^*=z_b' D_0/D(0)$ and $z_b^*=z_b D_0/D(0)$ are the extrapolation lengths at the input and output. Notice that this picture does not apply in real position space.

\section{Conclusion}
\label{sec:3}
We have reviewed the change of the distribution of transmission eigenvalues for diffusive samples when the effects of boundary reflectivities are included. A phase transition occurs at the point that the difference between the boundary reflectivities on the two ends of the sample $|z_{b}-z_{b}'|$ equals $L$ at which the singularity in $\rho(\cal T)$ as ${\cal T}\rightarrow 1$ changes from $(1-{\cal T})^{-\frac{1}{2}}$ to $(1-{\cal T})^{-\frac{1}{3}}$. For stronger asymmetry, perfect transmission cannot be achieved. However, the average conductance still obeys Ohm's law when boundary resistance terms are included. For diffusive systems, the average intensity profile decays linearly towards the output with the extrapolation length corresponding to the distance beyond the sample at which intensity extrapolates to zero. For localized waves, propagation can be described using a position-dependent diffusion coefficient given by Eq. (19) with $z_b'^*$ and $z_b^*$ the extrapolation lengths in virtual z-space related to position space via $dz=\frac{D(x)}{D_0}dx$. These results show that in addition to wavefront shaping, boundary reflectivity provides an additional degree of freedom to control the transmission and the energy distributions in the interior of disordered samples. 

\section*{Acknowledgements}

C. T. would like to thank the hospitality of Jean-Claude Garreau during his stay in Universit$\acute{\rm e}$ de Lille 1, where part of this work was done. This work is supported by the National Natural Science Foundation of China (Grant No. 11535011) and by National Science Foundation grant DMR/-BSF-1609218. A. Z. G. is grateful for many warm and stimulating discussions with Roger Maynard.


%


\begin{thebibliography}{}

\bibitem{Morse}
P. M. Morse, H. Feshbach, \textit{Methods of Theoretical Physics, Part II} (McGraw-Hill New York, 1953)

 
\bibitem{Lagendijk89}
A. Lagendijk, R. Vreeker,  P. De Vries, Phys. Lett. A \textbf{136}, (1989) 81-88

\bibitem{Pine91}
J. X. Zhu, D. J. Pine,  D. A. Weitz, Phys. Rev. A \textbf{44}, (1991) 3948-3959

\bibitem{Genack93}
J. H. Li, A. A. Lisyansky, T. D. Cheung, D. Livdan,  A. Z. Genack, Europhys. Lett. \textbf{22}, (1993) 675-680




\bibitem{1987a}
A. Z. Genack, Phys. Rev. Lett. \textbf{58},  (1987) 2043-2046

\bibitem{1997d}
D. S. Wiersma, P. Bartolini, A. Lagendijk,  R. Righini, Nature {\bf 390},  (1997) 671-673

\bibitem{Cheng13}
X. Cheng, C. Tian, A. Z. Genack, Phys. Rev. B \textbf{88}, (2013) 094202

\bibitem{Tian15}
L. Zhao, C. Tian, Y. P. Bliokh, V. Freilikher, Phys. Rev. B \textbf{92}, (2015) 094203


 
 
 
\bibitem{Dorokhov84}
O. N. Dorokhov, Solid State Commun. \textbf{51}, (1984) 381-384

\bibitem{Imry86}
Y. Imry, Euro. Phys. Lett. \textbf{1}, (1986) 249-256

\bibitem{Mello88}
P. A. Mello, P. Pereyra, N. Kumar, Ann. Phys. \textbf{181}, (1988) 290-317

\bibitem{Nazarov94}
Y. V. Nazarov, Phys. Rev. Lett. \textbf{73}, (1994) 134-137

\bibitem{Rossum95}
T. M. Nieuwenhuizen, M. C. W. van Rossum, Phys. Rev. Lett. \textbf{74}, (1995) 2674-2677

\bibitem{Kogan95}
E. Kogan, M. Kaveh, Phys. Rev. B, \textbf{52}, (1995) R3813-R3815

\bibitem{Beenakker97}
C. W. J. Beenakker, Rev. Mod. Phys. \textbf{69}, (1997) 731-808

\bibitem{Rossum99}
M. C. W. van Rossum, T. M. Nieuwenhuizen, Rev. Mod. Phys. \textbf{71}, (1999) 313-371

\bibitem{Abrahams79}
E. Abrahams, P. W. Anderson, D. C. Licciardello, T. V. Ramakrishnan, Phys. Rev. Lett. \textbf{42}, (1979) 673-676


\bibitem{1985g}
M. P. van Albada, A. Lagendijk, Phys. Rev. Lett. {\bf 55},  (1985) 2692

\bibitem{1985h}
P. Wolf, G. Maret, Phys. Rev. Lett. {\bf 55},  (1985) 2696



\bibitem{Maynard86}
E. Akkermans, P. E. Wolf, R. Maynard, Phys. Rev. Lett. \textbf{56}, (1986) 1471-1474

\bibitem{Maynard88}
E. Akkermans , P. E. Wolf, R. Maynard, G. Maret, J. Phys. France \textbf{49}, (1988) 77-98



\bibitem{Altshuler91}
B. L. Altshuler, P. A. Lee, R. A. Webb, \textit{Mesoscopic Phenomena in Solids} (Elsevier, 1991)

\bibitem{Akkermans07}
E. Akkermans, G. Montambaux, \textit{Mesoscopic Physics of Electrons and Photons} (Cambridge University, 2007)


\bibitem{Thouless77}
D. J. Thouless, Phys. Rev. Lett. \textbf{39}, (1977) 1167-1169


 
\bibitem{Anderson80}
P. W. Anderson, D. J. Thouless, E. Abrahams, D. S. Fisher, Phys. Rev. B \textbf{22}, (1980) 3519-3526

\bibitem{Fisher81}
D. S. Fisher, P. A. Lee, Phys. Rev. B \textbf{23}, (1981) 6851-6854

\bibitem{Genack14}
Z. Shi, J. Wang, A. Z. Genack, Proc. Nat. Acad. Sci. \textbf{111}, (2014) 2926-2930

\bibitem{Stone85}
P. A. Lee, A. D. Stone, Phys. Rev. Lett. \textbf{55}, (1985) 1622-1625


\bibitem{1990g} J.  L. Pichard, N. Zanon, Y. Imry, A. D. Stone, J. Phys. France {\bf 51}, (1990) 22 

\bibitem{Marko99}
P. Markoˇs, Phys. Rev. Lett. \textbf{83}, (1999) 588–591

\bibitem{Soukoulis01}
M. Ruhlander, C. M. Soukoulis,  Physica B \textbf{296}, (2001) 32–35

\bibitem{Gopar02}
V. Gopar, K. A. Muttalib, P. W¨olfle,  Phys. Rev. B \textbf{66}, (2002) 174-204


\bibitem{Mosk08}
I. M. Vellekoop, A. P. Mosk, Phys. Rev. Lett. \textbf{101}, (2008) 120601

\bibitem{Popoff10}
S. M. Popoff, G. Lerosey, R. Carminati, M. Fink, A. C. Boccara, S. Gigan, Phys. Rev. Lett. \textbf{104}, (2010) 100601

\bibitem{Choi11}
W. Choi, A. P. Mosk, Q. H. Park, W. Choi, Phys. Rev. B \textbf{83}, (2011) 134207

\bibitem{Popoff14}
S. Popoff, A. Goetschy, S. Liew, A. D. Stone,  H. Cao, Phys. Rev. Lett. \textbf{112}, (2014) 133903





\bibitem{Aubry14}
B. G$\acute{\rm e}$rardin, J. Laurent, A. Derode, C. Prada,  A. Aubry, Phys. Rev. Lett. \textbf{113}, (2014) 173901

\bibitem{2012f}
Z. Shi, A. Z. Genack, Phys. Rev. Lett. {\bf 108} (2012) 043901


\bibitem{Genack15}
M. Davy, Z. Shi, J. Park, C. Tian, A. Z. Genack, Nat. Commun. \textbf{6}, (2015) 7893

\bibitem{Mosk07}
I. M. Vellekoop, A. P. Mosk, Opt. Lett. \textbf{32}, (2007) 2309-2311

\bibitem{Mosk12}
A. P. Mosk, A. Lagendijk, G. Lerosey, M. Fink, Nat. Photon. \textbf{6}, (2012) 283-292

\bibitem{Cao16}
A. Yamilov, S. Petrenko, R. Sarma, H. Cao, Phys. Rev. B \textbf{93}, (2016) 100201





\bibitem{Dorokhov82}
O. N. Dorokhov,  Solid State Commun. \textbf{41}, (1982) 431-434

\bibitem{Efetov97}
K. Efetov, \textit{Supersymmetry in Disorder and Chaos} (Cambridge University Press, Cambridge, England, 1997)

\bibitem{Tian13}
C. Tian, Physica E \textbf{49}, (2013) 124-153


\bibitem{Tian08} C. Tian, Phys. Rev. B \textbf{77}, (2008) 064205



\bibitem{Tian10} C. S. Tian, S. K. Cheung, Z. Q. Zhang, Phys. Rev. Lett. \textbf{105}, (2010) 263905

\bibitem{Tian13a} L. Y. Zhao, C. S. Tian, Z. Q. Zhang, X. D. Zhang, Phys. Rev. B \textbf{88}, (2013) 155104



\bibitem{Wiersma00}
B. A. van Tiggelen, A. Lagendijk, D. S. Wiersma, Phys. Rev. Lett. \textbf{84}, (2000) 4333-4336


\end{thebibliography}
\end{document}